\documentclass[12pt,reqno]{amsart}

\usepackage{multirow}
\usepackage{hhline}

\usepackage{mathtools}
\usepackage{times}
\usepackage[T1]{fontenc}
\usepackage{mathrsfs}
\usepackage{latexsym}
\usepackage[dvips]{graphics}
\usepackage[titletoc, title]{appendix}
\usepackage{epsfig}
\usepackage{amssymb}
\usepackage{amsmath,amsfonts,amsthm,amssymb,amscd}
\usepackage{color}
\usepackage{hyperref}
\usepackage{url}
\usepackage{breakurl}
\newcommand{\bburl}[1]{\textcolor{blue}{\url{#1}}}
\usepackage{comment}

\newcommand{\be}{\begin{equation}}
\newcommand{\ee}{\end{equation}}
\newcommand{\seqnum}[1]{\href{https://urldefense.com/v3/__https://oeis.org/*1*7D*7B*5Crm__;IyUlJQ!!DZ3fjg!5yDnHeGSfGljEqbpuRw46hLMiV_jaL8rn2Yk7DqDl12SpLF1Wq4c5kckywJ55Db23mDqXkwmG3-qhLCpIIaQw7Y$   \underline{#1}}}

\newcommand{\rs}{{\rm RS}}
\newcommand{\ra}{{\rm RA}}

\newcommand{\RS}{{\rm RS}}
\newcommand{\RA}{{\rm RA}}
\newcommand{\Rave}{{\rm R_{{\rm ave}}}}

\newcommand{\R}{\ensuremath{\mathbb{R}}}
\newcommand{\C}{\ensuremath{\mathbb{C}}}

\newcommand{\E}{\mathbb{E}}

\numberwithin{equation}{section}

\begin{document}

\title[Applications of Improvements to the Pythagorean Won-Lost Expectation]{Applications of Improvements to the Pythagorean Won-Lost Expectation in Optimizing Rosters}

\author[Almeida]{Alexander F. Almeida} 
\email{\textcolor{blue}{\href{mailto:afa66@georgetown.edu}{afa66@georgetown.edu}}, 
\textcolor{blue}{\href{mailto:alexanderalmeida118@gmail.com}{alexanderalmeida118@gmail.com}}} 
\address{McDonough School of Business, Georgetown University, Washington D.C.}

\author[Dayaratna]{Kevin Dayaratna}
\email{\textcolor{blue}{\href{mailto:kd871@georgetown.edu}{kd871@georgetown.edu}},
\textcolor{blue}{\href{mailto:kdd0211@gmail.com}{kdd0211@gmail.com}}}
\address{Department of Mathematics and Statistics, Georgetown University, Washington DC}

\author[Miller]{Steven J. Miller}
\email{\textcolor{blue}{\href{mailto:sjm1@williams.edu}{sjm1@williams.edu}},
\textcolor{blue}{\href{mailto:Steven.Miller.MC.96@aya.yale.edu}{Steven.Miller.MC.96@aya.yale.edu}}}
\address{Department of Mathematics and Statistics, Williams College, Williamstown, MA 01267, USA}

\author[Yang]{Andrew K. Yang}
\email{\textcolor{blue}{\href{mailto:aky30@cam.ac.uk}{aky30@cam.ac.uk}},
\textcolor{blue}{\href{mailto:andrewkelvinyang@gmail.com}{andrewkelvinyang@gmail.com}}}
\address{Emmanuel College, University of Cambridge, Cambridge, CB2 3AP, UK}

\begin{abstract} Bill James' Pythagorean formula has for decades done an excellent job estimating a baseball team's winning percentage from very little data: if the average runs scored and allowed are denoted respectively by $\rs$ and $\ra$, there is some $\gamma$ such that the winning percentage is approximately $\rs^\gamma / (\rs^\gamma + \ra^\gamma)$. One important consequence is to determine the value of different players to the team, as it allows us to estimate how many more wins we would have given a fixed increase in run production. We summarize earlier work on the subject, and extend the earlier theoretical model of Miller (who estimated the run distributions as arising from independent Weibull distributions with the same shape parameter; this has been observed to describe the observed run data well). We now model runs scored and allowed as being drawn from independent Weibull distributions where the shape parameter is not necessarily the same, and then use the Method of Moments to solve a system of four equations in four unknowns. Doing so yields a predicted winning percentage that is consistently better than earlier models over the last 30 MLB seasons (1994 to 2023). This comes at a small cost as we no longer have a closed form expression but must evaluate a two-dimensional integral of two Weibull distributions and numerically estimate the solutions to the system of equations; as these are trivial to do with simple computational programs it is well worth adopting this framework and avoiding the issues of implementing the Method of Least Squares or the Method of Maximum Likelihood.
\end{abstract}



\thanks{We thank the editors and referees for comments that improved the analysis and exposition.}

\maketitle

\tableofcontents

%%%%%%%%%%%%%%%%%%%%%%%%%%%%%%%%%%%%%%%%%%%%%%%%%%%%%%%%%%%%%%%%%%%%%%%%%%%%%%%%%%%%%%%%%%%%%%%%%%%%%%%%%%%%%%%%%%%%%%%%%%%%%%%%%%%%%%%%%%%%%%%
%%%%%%%%%%%%%%%%%%%%%%%%%%%%%%%%%%%%%%%%%%%%%%%%%%%%%%%%%%%%%%%%%%%%%%%%%%%%%%%%%%%%%%%%%%%%%%%%%%%%%%%%%%%%%%%%%%%%%%%%%%%%%%%%%%%%%%%%%%%%%%%
%%%%%%%%%%%%%%%%%%%%%%%%%%%%%%%%%%%%%%%%%%%%%%%%%%%%%%%%%%%%%%%%%%%%%%%%%%%%%%%%%%%%%%%%%%%%%%%%%%%%%%%%%%%%%%%%%%%%%%%%%%%%%%%%%%%%%%%%%%%%%%%
\section{Introduction}

There are two related problems all baseball teams struggle to solve: win games (and championships) and make money. With finite resources, it is essential for teams to be efficient in determining whom to sign and for how much. It is a two step process to optimally make these decisions. First, teams must determine the net value of a player to their offense / defense. This can be done through metrics such as the ``runs created'' statistic \cite{A,C}, or more involved methods such as Monte Carlo simulation, which takes into account that players do not exist in a vacuum and one's contributions depends on the rest of the lineup.

To estimate how valuable a player that improves a team's offensive or defensive output is, we need to convert from runs scored and allowed to wins. In other words, how valuable is a given output? For example, see \cite{Ad, BZ, Br, Go, De, Ke, Min, Pe, Vo} for a sampling of how teams estimate performance and the impact this has on team construction, as well as recent work \cite{CJMMMPS} which shows that a natural extension of this formula leads to an excellent predictor of the probability one team beats another in the playoffs. This conversion has often been done by use of Bill James' Pythagorean Won-Lost formula \cite{Ja, Wi}, which for most teams leads to a simple rule of thumb that roughly every 10 net runs created translates to an additional win. It states that a team's winning percentage is well-estimated by \begin{equation}\label{eq:pythagWL}  \frac{\#{\rm Wins}}{\#{\rm Games}} \ \approx \ \frac{{\rm RS}^2}{{\rm RA}^2 + {\rm RS}^2}, \end{equation} where ${\rm RS}$ (resp. ${\rm RA}$) is the average number of runs scored (allowed) per game.\footnote{In ongoing work, Cleary, Jeffries, Miller, Miller, Murray, Pasha and Skiera \cite{CJMMMPS} adjust the Pythagorean formula to take into account the two teams playing by rescaling the runs scored and allowed relative to the league average. Thus if we have for the home team ${\rm RS}_{\rm h}, {\rm RA}_{\rm h}$ and for the away team ${\rm RS}_{\rm a}, {\rm RA}_{\rm a}$ and the league average runs scored per game is $R$, to calculate the probability the home team wins we use for runs scored ${\rm RS}_h ({\rm RA}_{\rm a} / {\rm R})$ and for runs allowed ${\rm RA}_h ({\rm RS}_{\rm a}/{\rm R})$; thus if the home team is playing a team that allows fewer runs than the league average, its run production is decreased. Looking at playoff data from 2001 through 2019, this modification predicted the home team should win 80.18 times and lose 68.82, essentially perfectly matching the observed 80 series victories and 69 series losses by the home team!} Though it is simple to compute with a standard calculator (or even with pen and paper) when the exponent is 2, simplicity such as this is not needed in the 21st century, and one can explore improvements. As teams are trying to optimize wins, revenue or both, the better they can predict the value of a player, the better they can solve these problems. Thus, rather than have an exponent of 2, sabermetricians explored, both numerically and theoretically, and found values of the exponent that do a better job. These values depend on the era and style of play: is it a pitcher's friendly environment, say the deadball era, or is it from a time when offensive production suddenly exploded?

We begin by summarizing earlier work on the subject. Our starting point is Miller's 2007 paper \cite{Mi}, where he showed that expressions of the form \eqref{eq:pythagWL} are consequences of reasonable models for run production; this advances the subject from experimental observations to a theoretical justification. The model makes several assumptions which range from the clearly false (runs scored and allowed are drawn from continuous and not discrete random variables) to the perhaps needlessly restrictive, perhaps not (specifically, both are modeled from three parameter Weibulls\footnote{A random variable $X$ follows a Weibull distribution with parameters $\alpha, \beta, \gamma$ if ${\rm Prob}(X \in [a,b]) = \int_a^b f(x;\alpha, \beta, \gamma)dx$, where $f(x;\alpha, \beta, \gamma) = (\gamma/\alpha) u^{\gamma - 1} \exp(-u^\gamma)$ with $u = (x - \beta)/\alpha$.} with the same shape parameter; see Appendix \ref{sec:weibull} for more details on this family of random variables). These assumptions are deliberately chosen to lead to a tractable mathematical model (see Appendix \ref{sec:millerRecap}, where we recall those arguments). While it does a very good job fitting the data, we present below a discussion of some previous improvements, followed by our new results and observations.

There are several earlier works worth noting. \\ \

\begin{itemize}

\item The reason Miller used the three parameter Weibull distribution is that the needed multivariable integral, namely the probability a team scores more runs than it allows, can be done in closed form (when the shape parameter $\gamma$ is the same for both), leading to \eqref{eq:pythagWL}. Luo and Miller \cite{LM} generalized to modeling the run distributions by linear combinations of Weibulls with the same shape parameter; interestingly, there is no significant improvement in predictive power. \\ \

\item Just as elections can be confidently called before all the results are in, so too can many games. Also in \cite{LM} the authors show that if we call games in late innings when one team is up by a lot, and adjust the runs scored and allowed averages for the team accordingly, there is no significant improvement in predictive power. In other words, the ``garbage'' runs scored or allowed when a team is winning or losing by a lot makes very little difference. Additionally, there was no noticeable gain in taking into account ballpark effects on run production.\\ \

\item While a simpler formula than \eqref{eq:pythagWL} is not needed, it is nice to have one that is easier for the average fan to use and understand, allowing a quick ballpark estimate for the value of decisions. This is similar to how many complicated statistics are often normalized and expressed in a way that is relatable to the general public. A linear version exists, originally observed by \cite{JT} and derived in \cite{CGLMP,DM} as a consequence of \eqref{eq:pythagWL} by doing a multivariate Taylor Series expansion. We reproduce that derivation in Appendix \ref{sec:linear}, and compare how easy it is to use and how well it predicts to other methods. \\ \

\end{itemize}

We then turn to our main contribution: exploring the potential improved predictive power when we allow more general distributions for runs scored and allowed. Miller's work led to determining the parameters of the Weibull distributions by approximating the observed distribution of runs scored or allowed by either the Method of Least Squares or the Method of Maximum Likelihood; while these are straightforward computations, it is a bit of a pain to code and use (though quite doable these days; unfortunately due to the intricate relationships there are no simple closed form solutions for the values that minimize the difference between predicted and observed run distributions).

Our main contribution here is to explore the consequences of no longer requiring a closed form expression for the integral of the probability the runs scored exceeds the runs allowed; it was that requirement that restricted earlier work to distributions such as the three parameter Weibull where both had the same shape parameter. In particular, we concentrate on the case when both runs scored and allowed are drawn from three parameter Weibulls, but we no longer require the shape parameter $\gamma$ to be the same for each (we do still take $\beta$ to be $-1/2$, as this has each bin centered about integer scores; thus the area under the curve from -1/2 to 1/2 corresponds to 0 runs, while 1/2 to 1 corresponds to 1 run and so on).\footnote{The advantage of this choice of $\beta$ is that the observed runs are never at the boundary of two bins.}

We now have four free shape parameters: $\alpha_{\rs}, \gamma_{\rs}, \alpha_{\ra}, \gamma_{\ra}$; we can numerically determine these by looking at the observed runs scored and allowed data and choosing the values of these parameters such that our continuous distributions have the same mean and variance as the data. While there are simple expressions for the mean and variance of a Weibull in terms of its parameters, the resulting system of equations cannot be solved in closed form, to say nothing about the subsequent problem of evaluating the multivariable integral which is the probability that the runs scored exceed the runs allowed; however, it is trivial with any reasonable modern computational system to immediately obtain excellent numerical approximations to the systems of equation and resulting integral.

We describe how to do this in \S\ref{sec:rsradifferentgammaweibull}, and report on the improvement this has in predictive ability by examining the last 30 years of MLB seasons (1994 to 2023). After this analysis, we use our approach to turn to the motivating question for this research: estimating the value of scoring additional runs given how many runs a team is scoring and allowing; our improvements in predictive power thus translate to better assessments of the worth of players. We then conclude in \S\ref{sec:future} with thoughts on future research.

%%%%%%%%%%%%%%%%%%%%%%%%%%%%%%%%%%%%%%%%%%%%%%%%%%%%%%%%%%%%%%%%%%%%%%%%%%%%%%%%%%%%%%%%%%%%%%%%%%%%%%%%%%%%%%%%%%%%%%%%%%%%%%%%%%%%%%%%%%%%%%%%%%%%%%%%%%%%%%%%%%
%%%%%%%%%%%%%%%%%%%%%%%%%%%%%%%%%%%%%%%%%%%%%%%%%%%%%%%%%%%%%%%%%%%%%%%%%%%%%%%%%%%%%%%%%%%%%%%%%%%%%%%%%%%%%%%%%%%%%%%%%%%%%%%%%%%%%%%%%%%%%%%%%%%%%%%%%%%%%%%%%%
%%%%%%%%%%%%%%%%%%%%%%%%%%%%%%%%%%%%%%%%%%%%%%%%%%%%%%%%%%%%%%%%%%%%%%%%%%%%%%%%%%%%%%%%%%%%%%%%%%%%%%%%%%%%%%%%%%%%%%%%%%%%%%%%%%%%%%%%%%%%%%%%%%%%%%%%%%%%%%%%%%
\section{Runs Scored and Allowed with Differently Shaped Weibulls}\label{sec:rsradifferentgammaweibull}

Work by Miller \cite{Mi}, and then extended by others both for baseball and other sports, established a statistical model that explicitly yields the Pythagorean Formula as a consequence of the assumptions: runs scored and allowed are independent random variables drawn from Weibulls with the same shape parameter. Our contribution is to remove the assumption that the shape parameter of the Weibull, $\gamma$, must be the same for both distributions. By introducing two different shape parameters, which we denote $\gamma_\rs$ and $\gamma_\ra$, we are able to obtain a better fit to the data and an improvement in predictive power, though at a cost: we no longer have a closed form expression for the winning percentage.

While of course runs are not drawn from continuous distributions, doing so leads to a tractable model that is quite close, year after year, to observed data. Further, as remarked earlier, by setting the shift parameter $\beta$ to be $-1/2$ we remove all edge effects from the discreteness of the observed run distributions, with those values now separated as the centers of our bins. Miller introduced the Method of Least Squares, which fits the values of $\alpha_\rs$, $\alpha_\ra$ and $\gamma$ by minimizing the squared error between the Weibull distribution and the bins of the observed, discrete run distribution (see \cite{Mi}, Section 3, footnote 2). Miller showed that the Method of Least Squares gave similarly good win predictions in the 2004 American League season than those obtained through estimating $\alpha_\rs$, $\alpha_\ra$ and $\gamma$ by the Method of Maximum Likelihood. We choose to use the Method of Least Squares, and we update it by now allowing $\gamma_\rs$ and $\gamma_\ra$ to be different, leading to better fits and improved predictive power.

We now introduce the Method of Moments. Instead of finding the values of the parameters that lead to minimizing errors with the observed run histograms, we find the four values by setting the means and variances equal. This leads to a significantly easier method to implement than the earlier works, which proceeded by varying parameters in applications of the Method of Least Squares or the Method of Maximum Likelihood to find the optimal values; now we just numerically approximate the solution to two different systems of two equations with two unknowns, and then estimate the resulting two-dimensional integral. The new predictive value is marginally better than the Pythagorean formula with shape parameter 1.83, which is used by Baseball Reference and believed to be the value with the strongest predictive power \cite{Ba}. Hereafter we refer to the prediction from James' formula with exponent 1.83 as Pythag(1.83).

The Method of Moments often outperforms even the updated Method of Least Squares, despite requiring considerably less data and computation. It narrowly outperforms Pythag(1.83), and hugely outperforms the old Method of Least Squares in \cite{Mi}. This leap should be expected as a team likely does not score and allow runs under the same shaped distributions. While forcing the two shape parameters to be equal results in easier integrals which can be done in closed form, it loses important information. While this model improves accuracy, it loses closed form results.

%%%%%%%%%%%%%%%%%%%%%%%%%%%%%%%%%%%%%%%%%%%%%%%%%%%%%%%%%%%%%%%%%%%%%%%%%%%%%%%%
%%%%%%%%%%%%%%%%%%%%%%%%%%%%%%%%%%%%%%%%%%%%%%%%%%%%%%%%%%%%%%%%%%%%%%%%%%%%%%%%
%%%%%%%%%%%%%%%%%%%%%%%%%%%%%%%%%%%%%%%%%%%%%%%%%%%%%%%%%%%%%%%%%%%%%%%%%%%%%%%%
\subsection{Method of Moments}

As remarked, the probability density function of the Weibull distribution is
\begin{equation}
    f(x;\alpha,\beta,\gamma) \ := \ \frac{\gamma}{\alpha}((x-\beta)/\alpha)^{\gamma -1}e^{-((x-\beta)/\alpha)^\gamma}
\end{equation}
for $x \geq \beta$; $\alpha$, $\beta$ and $\gamma$ are the three parameters of the distribution. While we are able to model a variety of curves by appropriately choosing values for these parameters, the possibilities are not as extensive as one might think, as $\alpha$ and $\beta$ just respectively rescale and translate the distribution; it is only $\gamma$ that changes the shape.

\begin{figure}[h]
\begin{center}
\scalebox{.9125}{\includegraphics{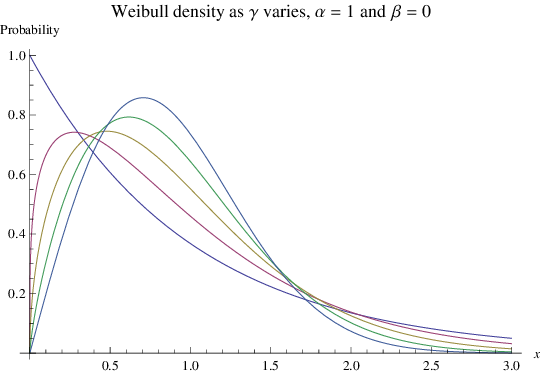}}
\caption{\label{fig:pythagweibullgamma1to4} The changing probabilities of a family of Weibulls with $\alpha = 1$, $\beta = 0$, and $\gamma \in \{1, 1.25, 1.5, 1.75, 2\}$; $\gamma = 1$ corresponds to the exponential distribution, and increasing $\gamma$ results in the bump moving rightward.}
\end{center}
\end{figure}

We illustrate the effect of different choices of $\gamma$ in Figure \ref{fig:pythagweibullgamma1to4} (taken from \cite{CGLMP}). As $\alpha$ and $\beta$ just rescale and translate, without loss of generality we set their values to be 1 and 0.

\ \\

We now describe how to use the Method of Moments to estimate the winning percentage. Let the runs scored and runs allowed per game be drawn from independent Weibull distributions with parameters ($\alpha_\rs, \beta = -1/2, \gamma_\rs$) and ($\alpha_\ra, \beta = -1/2, \gamma_\ra$), respectively. Straightforward integration yields closed form expressions for the mean $\mu_{\alpha,\beta,\gamma}$ and the variance $\sigma^2_{\alpha,\beta,\gamma}$ of a Weibull distribution in terms of its parameters and the Gamma function\footnote{Though we do not need this result, it is worth noting that the Gamma function generalizes the factorial function: $\Gamma(n+1) = n!$ when $n$ is a non-negative integer}, $\Gamma(s)$:
\begin{equation}\label{eq:gammadefn} \Gamma(s) \ := \ \int_0^\infty e^{-x} x^{s-1} dx,  \ \ \ {\rm Re(s) > 0}. \end{equation}

After straightforward integration, we find
\begin{eqnarray}\label{eq:meanvarweibull}
\mu_{\alpha,\beta,\gamma} &\ =\ & \alpha \Gamma\left(1+\gamma^{-1}\right) + \beta \nonumber\\ \sigma^2_{\alpha,\beta,\gamma}&\ =\ & \alpha^2 \Gamma\left(1+2\gamma^{-1}\right)  - \alpha^2 \Gamma\left(1+\gamma^{-1}\right)^2;
\end{eqnarray} for a derivation, see Appendix \ref{sec:weibull}. Note \eqref{eq:meanvarweibull} gives us two equations with two unknowns. We thus expect a solution to exist; while we cannot find a closed form expression for the parameters in terms of the observed mean and variance, we can easily approximate these values.

Let $\widehat{\mu}_\rs$ be an estimate for a team's mean runs scored per game, $\widehat{\sigma_\rs^2}$ an estimate for the variance in runs scored per game, $\widehat{\mu}_\ra$ an estimate for the mean runs allowed per game, and $\widehat{\sigma_{\ra}^2}$ an estimate for the variance in runs allowed per game. In our investigations, these are the sample means and sample variances of a team's runs scored and runs allowed per game over the course of the 2022 season.

We can now solve for $\alpha_\rs$, $\alpha_\ra$, $\gamma_\rs$ and $\gamma_\ra$ in the following system of equations (gradient descent and grid search both work well and efficiently):
\begin{eqnarray}
\label{meanvarianceeqns}
\widehat{\mu}_{\rs} &\ =\ & \alpha_\rs \Gamma\left(1+\gamma_\rs^{-1}\right) + \beta \nonumber\\ \widehat{\sigma_{\rs}^2} &\ =\ & \alpha_\rs^2 \Gamma\left(1+2\gamma_\rs^{-1}\right)  - \alpha_\rs^2 \Gamma\left(1+\gamma_{\rs}^{-1}\right)^2 \nonumber\\ \widehat{\mu}_{\ra} &\ =\ & \alpha_\ra \Gamma\left(1+\gamma_\ra^{-1}\right) + \beta \nonumber\\ \widehat{\sigma_{\ra}^2}&\ =\ & \alpha_\ra^2 \Gamma\left(1+2\gamma_\ra^{-1}\right)  - \alpha_\ra^2 \Gamma\left(1+\gamma_\ra^{-1}\right)^2.
\end{eqnarray}

Let $X$ and $Y$ be random variables modeling respectively the runs scored and runs allowed per game, drawn from independent Weibull distributions with parameters ($\alpha_\rs$, $-1/2$, $\gamma_\rs$) and ($\alpha_\ra$, $-1/2$, $\gamma_\ra$). Then the winning percentage is
\begin{eqnarray}
\label{diffgammawinpct}
& & \mbox{Prob}(X > Y) \ = \ \int_{x=\beta}^\infty \int_{y=\beta}^x f(x;\alpha_\rs,\beta,\gamma_\rs) f(y;\alpha_\ra,\beta,\gamma_\ra) {\rm d} y\; {\rm d} x \nonumber\\ & & = \
\int_{x=0}^\infty\tfrac{\gamma_\rs}{\alpha_\rs} \left(\tfrac{x}{\alpha_\rs}\right)^{\gamma_\rs-1} \exp\left(-\left(\tfrac{x}{\alpha_\rs}\right)^{\gamma_\rs}\right)\nonumber\\ & & \ \ \ \cdot \left[
\int_{y=0}^{x} \tfrac{\gamma_\ra}{\alpha_\ra}\left(\tfrac{y}{\alpha_{\ra}}\right)^{\gamma_\ra-1} \exp\left(-\left(\tfrac{x}{\alpha_\ra}\right)^{\gamma_\ra}\right) {\rm d} y \right] {\rm d} x  \nonumber\\ & & = \ \int_{x=0}^\infty\frac{\gamma_\rs}{\alpha_\rs}
\left(\frac{x}{\alpha_\rs}\right)^{\gamma_\rs-1} \exp\left(-(x/\alpha_\rs)^{\gamma_\rs}\right) \left[1
- \exp\left(-(x/\alpha_{\ra})^{\gamma_\ra}\right)\right] {\rm d} x  \nonumber\\ & & = \ 1 -
\int_{x=0}^\infty\frac{\gamma_\rs}{\alpha_\rs}
\left(\frac{x}{\alpha_\rs}\right)^{\gamma_\rs-1} \exp\left[-(x/\alpha_\rs)^{\gamma_\rs} -(x/\alpha_{\ra})^{\gamma_\ra}\right] {\rm d} x.
\end{eqnarray} If the shape exponents are the same, a simple change of variables leads to a closed form expression for the above (see \cite{CGLMP, Mi}; for completeness we reproduce this derivation in Appendix \ref{sec:millerRecap}); while we are not so fortunate in this more general setting, the resulting integral can be quickly computed numerically with high accuracy by Riemann sums, or better yet Simpson's Rule. Other good numerical methods include quadrature or Monte Carlo methods.

%%%%%%%%%%%%%%%%%%%%%%%%%%%%%%%%%%%%%%%%%%%%%%%%%%%%%%%%%%%%%%%%%%%%%%%%%%%%%%%%
%%%%%%%%%%%%%%%%%%%%%%%%%%%%%%%%%%%%%%%%%%%%%%%%%%%%%%%%%%%%%%%%%%%%%%%%%%%%%%%%
%%%%%%%%%%%%%%%%%%%%%%%%%%%%%%%%%%%%%%%%%%%%%%%%%%%%%%%%%%%%%%%%%%%%%%%%%%%%%%%%
\subsection{Analysis}

We use the Method of Moments to analyse the 30 teams, which are ordered by the number of overall season wins, from the 2022 season to see how closely our model fits the observed scoring patterns. For each team we compute the sample mean runs scored and allowed per game, and the sample variance in runs scored and allowed per game. We solve for the $\alpha_\rs$, $\alpha_\ra$, $\gamma_\rs$ and $\gamma_\ra$ that satisfy \eqref{meanvarianceeqns}, and then compute the win percentages by \eqref{diffgammawinpct}.

%%%%%% METHOD OF MOMENTS PREDICTIONS GRAPHIC HERE%%%%%%
\begin{table}[h]
\begin{center}
\resizebox{\textwidth}{!}{%
\begin{tabular}{lrrrrrrrrrrrrrr}
\hline
 {\rm Team} & \ &   {\rm Obs} {\rm W} & \ &   {\rm Pred} {\rm W} & \ &   {\rm Obs \%} & \ &   {\rm Pred \%} & \ &   {\rm Diff W} & \ &   {\rm      $\gamma_\rs $  } & \ &   {\rm      $\gamma_\ra $  } \\ \hline
 {\rm Los Angeles Dodgers} & \ &   111 & \ &   113.3 & \ &   0.685 & \ &   0.699 & \ &   -2.3 & \ &   1.88 & \ & 1.55 \\
 {\rm Houston Astros} & \ &   106 & \ &   100.4 & \ &   0.654 & \ &   0.620 & \ &   5.6 & \ &   1.56 & \ & 1.55 \\
 {\rm Atlanta Braves} & \ &   101 & \ &   98.8 & \ &   0.623 & \ &   0.610 & \ &   2.2 & \ &   1.80 & \ & 1.55 \\
 {\rm New York Mets} & \ &   101 & \ &   97.4 & \ &   0.623 & \ &   0.601 & \ &   3.6 & \ &   1.74 & \ & 1.51 \\
 {\rm New York Yankees} & \ &   99 & \ &   98.1 & \ &   0.611 & \ &   0.605 & \ &   0.9 & \ &   1.47 & \ & 1.70 \\
 {\rm St. Louis Cardinals} & \ &   93 & \ &   91.0 & \ &   0.574 & \ &   0.562 & \ &   2.0 & \ &   1.51 & \ & 1.57 \\
 {\rm Cleveland Guardians} & \ &   92 & \ &   85.4 & \ &   0.568 & \ &   0.527 & \ &   6.6 & \ &   1.60 & \ & 1.74 \\
 {\rm Toronto Blue Jays} & \ &   92 & \ &   88.5 & \ &   0.568 & \ &   0.546 & \ &   3.5 & \ &   1.58 & \ & 1.59 \\
 {\rm Seattle Mariners} & \ &   90 & \ &   87.5 & \ &   0.556 & \ &   0.540 & \ &   2.5 & \ &   1.70 & \ & 1.66 \\
 {\rm San Diego Padres} & \ &   89 & \ &   83.6 & \ &   0.549 & \ &   0.516 & \ &   5.4 & \ &   1.46 & \ & 1.55 \\
 {\rm Philadelphia Phillies} & \ &   87 & \ &   88.0 & \ &   0.537 & \ &   0.543 & \ &   -1.0 & \ &   1.58 & \ & 1.39 \\
 {\rm Milwaukee Brewers} & \ &   86 & \ &   83.9 & \ &   0.531 & \ &   0.518 & \ &   2.1 & \ &   1.64 & \ & 1.66 \\
 {\rm Tampa Bay Rays} & \ &   86 & \ &   86.5 & \ &   0.531 & \ &   0.534 & \ &   -0.5 & \ &   1.70 & \ & 1.63 \\
 {\rm Baltimore Orioles} & \ &   83 & \ &   79.9 & \ &   0.512 & \ &   0.493 & \ &   3.1 & \ &   1.59 & \ & 1.58 \\
 {\rm Chicago White Sox} & \ &   81 & \ &   81.4 & \ &   0.500 & \ &   0.503 & \ &   -0.4 & \ &   1.64 & \ & 1.40 \\
 {\rm San Francisco Giants} & \ &   81 & \ &   82.0 & \ &   0.500 & \ &   0.506 & \ &   -1.0 & \ &   1.58 & \ & 1.63 \\
 {\rm Boston Red Sox} & \ &   78 & \ &   79.4 & \ &   0.481 & \ &   0.490 & \ &   -1.4 & \ &   1.60 & \ & 1.42 \\
 {\rm Minnesota Twins} & \ &   78 & \ &   82.4 & \ &   0.481 & \ &   0.509 & \ &   -4.4 & \ &   1.64 & \ & 1.60 \\
 {\rm Arizona Diamondbacks} & \ &   74 & \ &   78.9 & \ &   0.457 & \ &   0.487 & \ &   -4.9 & \ &   1.68 & \ & 1.58 \\
 {\rm Chicago Cubs} & \ &   74 & \ &   75.1 & \ &   0.457 & \ &   0.464 & \ &   -1.1 & \ &   1.45 & \ & 1.47 \\
 {\rm Los Angeles Angels} & \ &   73 & \ &   78.0 & \ &   0.451 & \ &   0.482 & \ &   -5.0 & \ &   1.59 & \ & 1.51 \\
 {\rm Miami Marlins} & \ &   69 & \ &   71.1 & \ &   0.426 & \ &   0.439 & \ &   -2.1 & \ &   1.47 & \ & 1.64 \\
 {\rm Colorado Rockies} & \ &   68 & \ &   65.7 & \ &   0.420 & \ &   0.406 & \ &   2.3 & \ &   1.57 & \ & 1.76 \\
 {\rm Texas Rangers} & \ &   68 & \ &   76.9 & \ &   0.420 & \ &   0.475 & \ &   -8.9 & \ &   1.70 & \ & 1.81 \\
 {\rm Detroit Tigers} & \ &   66 & \ &   63.1 & \ &   0.407 & \ &   0.390 & \ &   2.9 & \ &   1.57 & \ & 1.76 \\
 {\rm Kansas City Royals} & \ &   65 & \ &   66.8 & \ &   0.401 & \ &   0.413 & \ &   -1.8 & \ &   1.57 & \ & 1.63 \\
 {\rm Cincinnati Reds} & \ &   62 & \ &   65.6 & \ &   0.383 & \ &   0.405 & \ &   -3.6 & \ &   1.53 & \ & 1.72 \\
 {\rm Pittsburgh Pirates} & \ &   62 & \ &   63.7 & \ &   0.383 & \ &   0.393 & \ &   -1.7 & \ &   1.61 & \ & 1.53 \\
 {\rm Oakland Athletics} & \ &   60 & \ &   61.7 & \ &   0.370 & \ &   0.381 & \ &   -1.7 & \ &   1.46 & \ & 1.68 \\
 {\rm Washington Nationals} & \ &   55 & \ &   55.5 & \ &   0.340 & \ &   0.342 & \ &   -0.5 & \ &   1.44 & \ & 1.97 \\
\hline
\end{tabular}%
}
\caption{\label{table:pythagmoments} Results from the Method of Moments, displaying the observed and predicted number of wins, winning percentage, and difference in games won and predicted for the 2022 season.}
\end{center}
\end{table}

In Table \ref{table:pythagmoments} we find that indeed many teams have large differences between their $\gamma_\rs$ and $\gamma_\ra$ values. However, every run scored for one team is a run allowed by another, so we expect the league average values of $\gamma_\rs$ and $\gamma_\ra$ to be similar. Indeed, we find that the league average of $\gamma_\rs$ is 1.59, while the league average of $\gamma_\ra$ is 1.61.

Across the league in 2022, the mean difference between predicted and observed wins is -0.01, suggesting the Method of Moments is unbiased, but as the difference can take negative values, that alone is no cause for celebration. A statistic more indicative of predictive power to consider is the absolute difference between predicted and observed wins for each team. In Appendix \ref{sec:linear}, see Table \ref{table:comparepredict} for a comparison of the absolute differences of different methods in the 2012 and 2022 seasons.

\begin{figure}[h]
\begin{center}
\scalebox{.9125}{\includegraphics{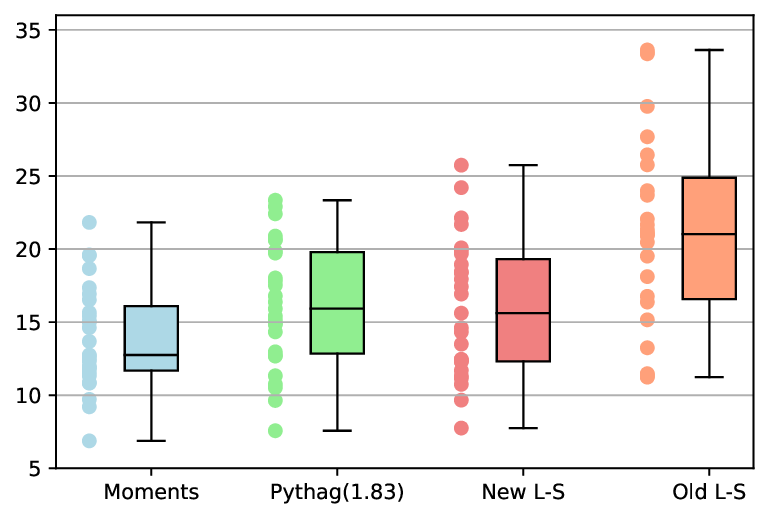}}
\caption{\label{fig:comparepredict} Scatter plot with boxplot representation for each season of the last 30 years (excl. 1994, 1995, 2020) of the Mean Squared Error in Predicted vs. Observed wins yielded by the four different methods: Moments, Pythag(1.83), ``New'' Least Squares ($\gamma_\rs,\gamma_\ra$ free), and ``Ol'' Least Squares ($\gamma_\rs=\gamma_\ra$). Tied games were included in the data, and counted as 0.5 observed wins for both teams.}
\end{center}
\end{figure}

To further justify the merits of the Method of Moments, we repeated this analysis over all 30 seasons since 1994. A common and useful statistic to evaluate estimators is the \emph{Mean Squared Error}, since that accounts for both bias and variance. After finding the mean squared error between predicted and observed wins for each method and team over the last 30 years of data, we find empirical evidence that the Method of Moments outperforms Pythag(1.83) and the Method of Least Squares. We present a boxplot illustrating these results in Figure \ref{fig:comparepredict}. The interquartile range of the boxplot clearly exhibits the superiority of the Method of Moments over the last 30 years.

%%%%%% COMPARE METHODS HERE %%%%%%
%%%%

It is clear that the old Method of Least Squares with $\gamma_\rs=\gamma_\ra$ typically has a larger mean squared error and performs the poorest. Even though it is using the amount of runs observed in \emph{every} game of the season to fit the Weibulls, rather than just the sample first and second moments, we see that forcing $\gamma_\rs=\gamma_\ra$ leads to worse fits and worse results.

\begin{figure}[h]
\begin{center}
\scalebox{.57}{\includegraphics{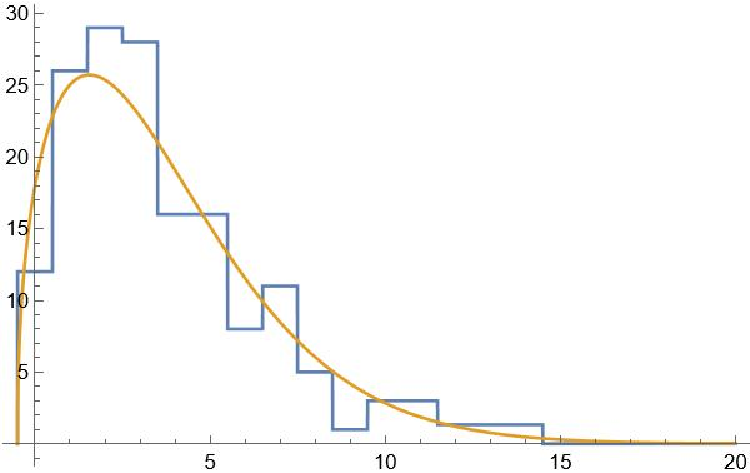}}\ \scalebox{.57}{\includegraphics{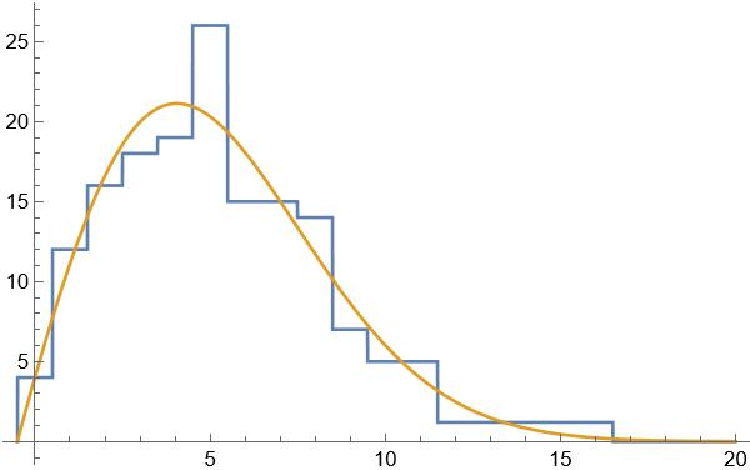}}
\caption{\label{figure:WSNmoments} For the 2022 Washington Nationals, comparison of the Weibulls produced by the \emph{Method of Moments} against the observed distribution of runs scored (left) and runs allowed (right) per game.}
\end{center}
\end{figure}

\begin{figure}[h]
\begin{center}
\scalebox{.57}{\includegraphics{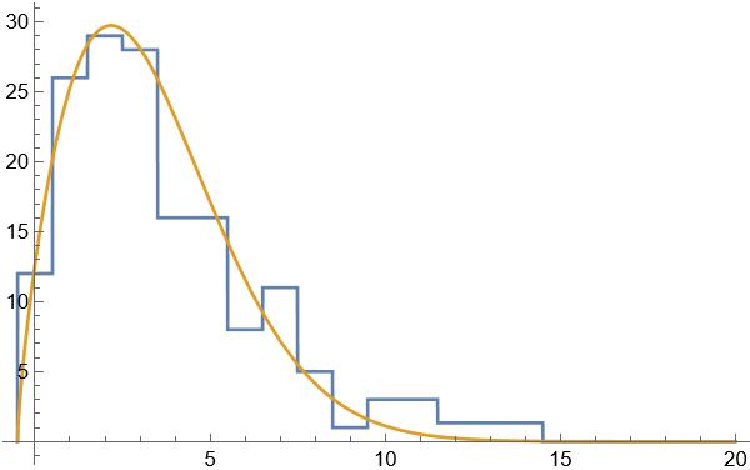}}\ \scalebox{.57}{\includegraphics{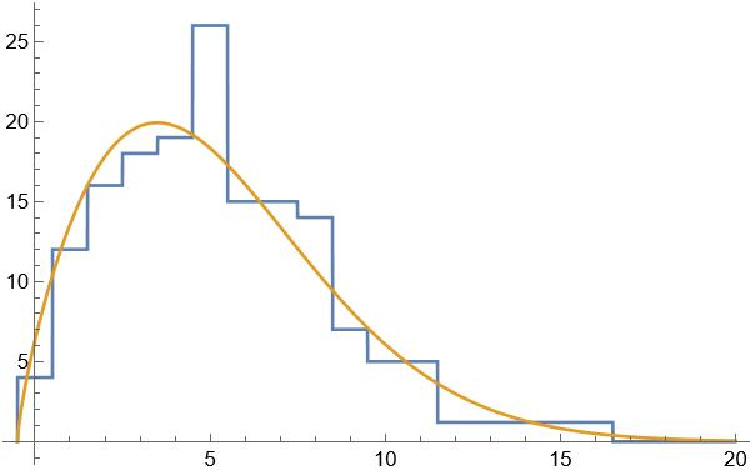}}
\caption{\label{figure:WSNleastsquares} For the 2022 Washington Nationals, comparison of the Weibulls produced by the \emph{Method of Least Squares} against the observed distribution of runs scored (left) and runs allowed (right) per game.}
\end{center}
\end{figure}

We illustrate this in Figure \ref{figure:WSNmoments}, showing the observed run distribution for the Washington Nationals in the 2022 season -- the team with the largest difference between $\gamma_\rs$ and $\gamma_\ra$ -- against the Weibulls produced by the Method of Moments. The runs scored data is heavily packed around 0 to 3 runs, while the runs allowed data is comparatively spread. The flexibility in shape allows the Weibulls to capture this. Compare how well these fit to Figure \ref{figure:WSNleastsquares}, with the Weibulls produced by the Method of Least Squares. Even though the Weibulls are fitted as closely as possible to the observed data, the restriction that both distributions have the same $\gamma$ still results in a slightly weaker fit.

%%%%%%%%%%%% FIGURE WSN 2022 %%%%%%%%%%%%%%%%%%%%%%%%%%%%%%
%%%%%%
%%%%%%%%%%%%%%%%

It is interesting that even when we allow $\gamma_\rs, \gamma_\ra$ to be free, the Method of Least Squares still performs marginally worse than the less computationally intensive Method of Moments and Pythag(1.83). This may due to \emph{overfitting}. By only using the sample means and variances to fit the Weibulls, we have lost information by processing the data, but have enough information to make a reasonable fit, and are more robust against data that may be misleading than an approach that uses every bit of information possible.

Finally, observe that Pythag(1.83) is actually using the first moment, since total runs scored/allowed is just the average runs scored/allowed multiplied by a constant number of games played. It does not matter whether we use the average or total runs scored/allowed in the Pythagorean formula. What we have shown is that by introducing just the second moment, our Method of Moments not only outperforms Pythag(1.83), but does so with theoretical backing from our statistical model.

We have made the CSV files of the last 30 years of raw data and the Jupyter Notebook files used to produce these findings publicly available at \cite{YM} for easy reproduction. The data was originally sourced from \cite{Ba}.

%%%%%%%%%%%%%%%%%%%%%%%%%%%%%%%%%%%%%%%%%%%%%%%%%%%%%%%%%%%%%%%%%%%%%%%%%%%%%%%%%%%%%%%%%%%%%%%%%%%%%%%%%%%%%%%%%%%%%%%%%%%%%%%%%%%%%%%%%%%%%%%%%%%%%%%%%%%%%%%%%%%%%%%%
%%%%%%%%%%%%%%%%%%%%%%%%%%%%%%%%%%%%%%%%%%%%%%%%%%%%%%%%%%%%%%%%%%%%%%%%%%%%%%%%%%%%%%%%%%%%%%%%%%%%%%%%%%%%%%%%%%%%%%%%%%%%%%%%%%%%%%%%%%%%%%%%%%%%%%%%%%%%%%%%%%%%%%%%
%%%%%%%%%%%%%%%%%%%%%%%%%%%%%%%%%%%%%%%%%%%%%%%%%%%%%%%%%%%%%%%%%%%%%%%%%%%%%%%%%%%%%%%%%%%%%%%%%%%%%%%%%%%%%%%%%%%%%%%%%%%%%%%%%%%%%%%%%%%%%%%%%%%%%%%%%%%%%%%%%%%%%%%%
\section{The Pythagorean Formula: Applications}\label{sec:applications}

In this section we apply the Pythagorean Formula to a critical economic problem for a baseball team - valuing players. We perform a similar analysis as in Section 5 of \cite{CGLMP} to estimate the value a given player brings to their team. Note the answer depends on how many runs the team scores and allows; not surprisingly adding 50 runs to a team that scores few might be significantly more valuable than adding 50 runs to an already productive team.

Specifically, if our team scores $x$ runs and allows $y$ across a season, how much should we pay to sign someone whom we estimate would increase our run production by $s$? For now we will focus only on how many additional wins they generate, treating all wins equally; this of course is a false assumption, as not all extra wins are created equal. Going from 55 to 65 wins in a season doesn't alter the fact that the season was a bad one, but going from 85 wins to 95 wins is often the difference between making the playoffs or not!

We proceed with the staple model Pythag(1.83), since this formula is not only the most robust, but also requires very little data - only the total or average runs scored and allowed. Once we estimate the amount of runs a player would contribute to our team, we can immediately compute the change in predicted wins. In \S\ref{sec:future}, we discuss the possibility of incorporating variance in runs scored and allowed into player analysis.

\begin{figure}[h]
\begin{center}
%%%\scalebox{.5}{\includegraphics{aky30contour1.jpg}}\ \scalebox{.5}{\includegraphics{aky30contour2.jpg}}
\scalebox{.4}{\includegraphics{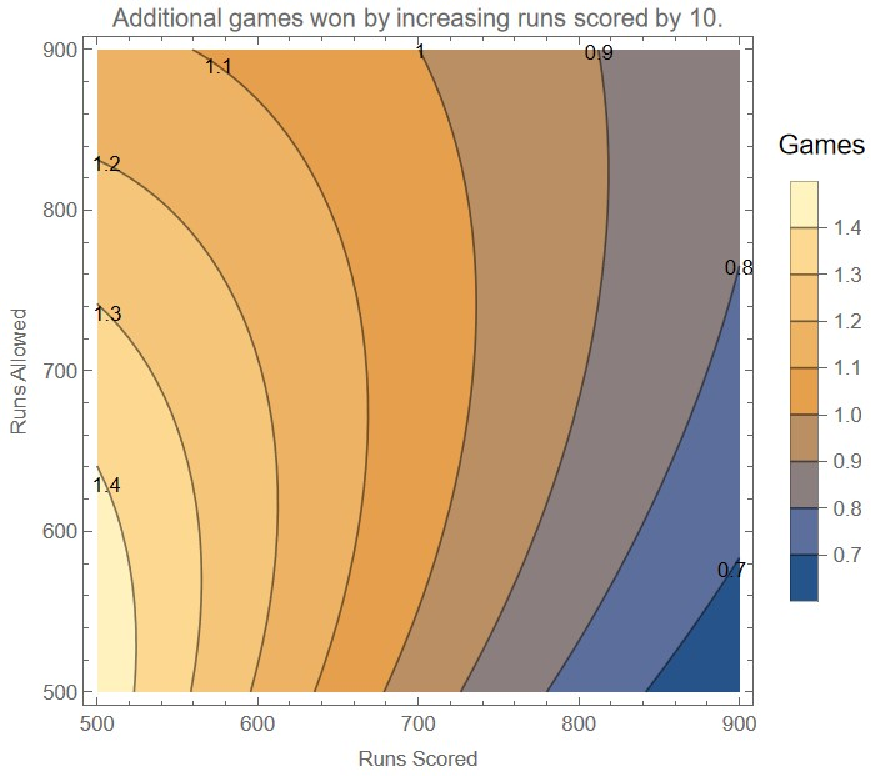}}\ \scalebox{.4}{\includegraphics{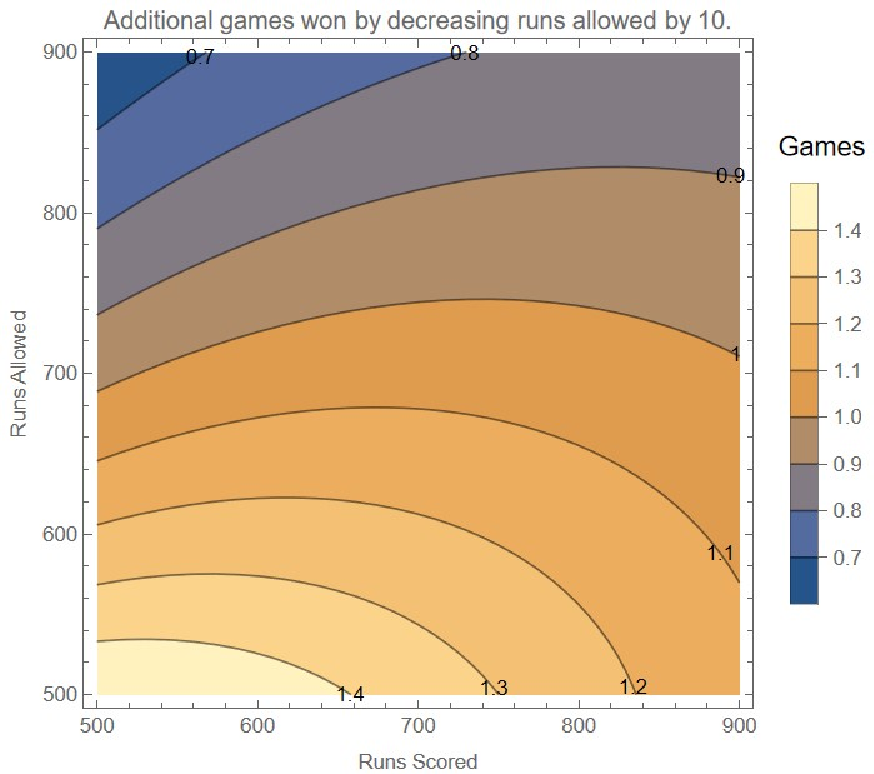}}
\caption{\label{fig:pythaggainsave} The predicted number of additional wins under Pythag(1.83) when: (left) scoring 10 more per season; (right) preventing 10 more per season.}
\end{center}
\end{figure}

In Figure \ref{fig:pythaggainsave}, we consider a range of runs scored and allowed per season that a team may currently operate at, and plot the additional wins per season that both a player who adds 10 runs a season is expected to give that team, and similarly for a player who saves 10 runs a season. We plot in the ballpark of 700 runs scored per season, which is close to the average for most MLB seasons (including 2022, at 693 runs). We deliberately chose $s=10$, as the common adage goes ``every 10 additional runs translates to one more win per season'' (see \cite{Bi}).

\begin{figure}[h]
\begin{center}
%%%\scalebox{.5}{\includegraphics{aky30contour3.jpg}}\ \scalebox{.5}{\includegraphics{aky30contour4.jpg}}
\scalebox{.4}{\includegraphics{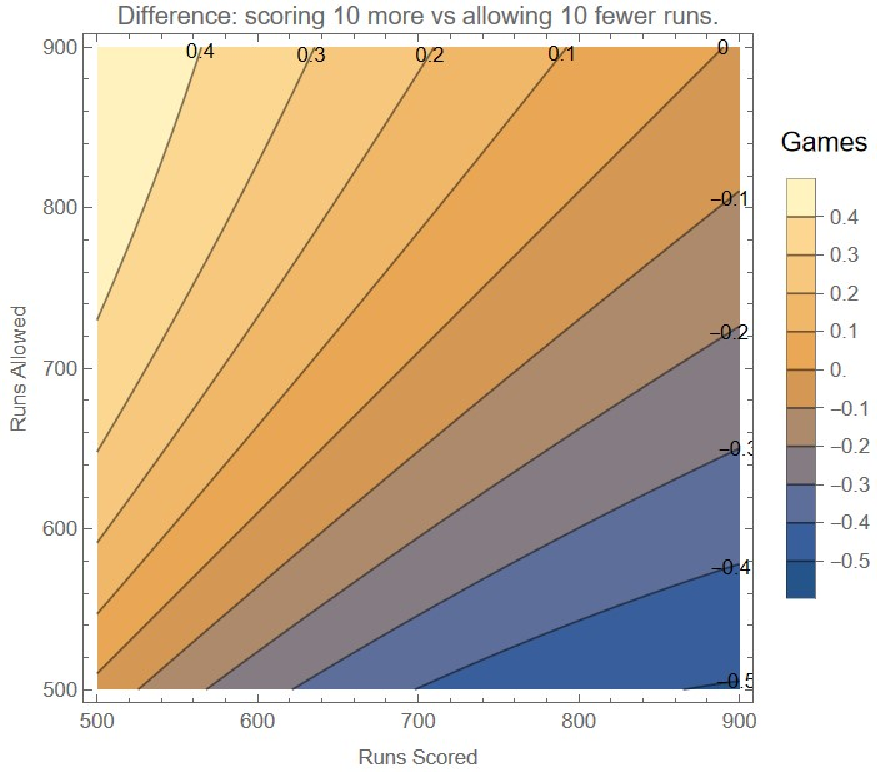}}\ \scalebox{.6}{\includegraphics{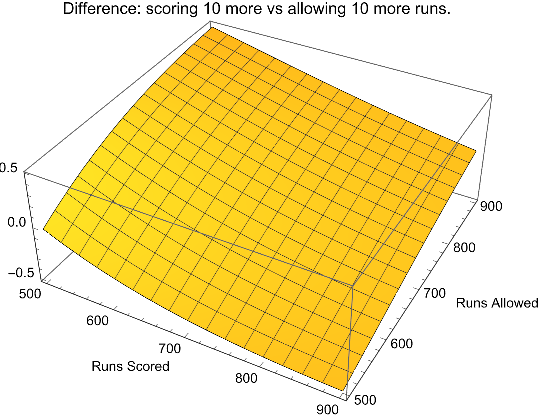}}
\caption{\label{fig:pythaggainminussave} The difference in the predicted number of additional wins under Pythag(1.83) from scoring 10 more per season versus allowing 10 fewer per season.}
\end{center}
\end{figure}

The Pythagorean formula allows us to quantify the value of scoring or preventing runs; see Figure \ref{fig:pythaggainminussave}, where we plot the difference in wins gained from scoring 10 more runs to wins gained from allowing 10 fewer runs.

While half a win may not sound like much, for the most competitive teams, any edge could be decisive. All teams make hundreds of these decisions every season, and the best teams get them right more often. Across the course of several seasons, both the money saved from better player evaluations, or the wins earned from wiser purchases, could very well provide a winning edge in a sport of extremely fine margins.

\subsection{Applications to Other Sports}

The Pythagorean Formula is flexible, and modelling win percentage from runs/points/goals scored and allowed can be transferred effectively to different sports (see \cite{Min}) by choosing a suitable exponent $\gamma$. The formula works better for high-scoring sports like basketball than for low-scoring sports like soccer. This is because an average soccer game only has around three goals scored in total \cite{Fo}, meaning very often a single important goal can decide a match, while in basketball a single successful shot holds much less significance over the result of the match. This additional randomness translates to more variance in the performance of Pythagorean expectation.

However, the main issue with soccer in particular is the prevalence of draws, which make up around 20\% \cite{Fo} of results in the English Premier League. A common strategy for mid-table teams is to play very defensively and hardly attempt to score against better opposition, settling for a draw. Taking draws as being worth half a win to each team would work well with the Pythagorean formula, but is not how the soccer league table works - a win is worth \emph{three} \emph{points}, and a draw is worth just a single point. Thus a different approach is needed to model the relationship between goals and points, and points are how teams are ranked at the end of the season.

%%%%%%%%%%%%%%%%%%%%%%%%%%%%%%%%%%%%%%%%%%%%%%%%%%%%%%%%%%%%%%%%%%%%%%%%%%%%%%%%%%%%%%%%%%%%%%%%%%%%%%%%%%%%%%%%%%%%%%%%%%%%%%%%%%%%%%%%%%%%%%%%%%%%%%%%%%%%%%%%
%%%%%%%%%%%%%%%%%%%%%%%%%%%%%%%%%%%%%%%%%%%%%%%%%%%%%%%%%%%%%%%%%%%%%%%%%%%%%%%%%%%%%%%%%%%%%%%%%%%%%%%%%%%%%%%%%%%%%%%%%%%%%%%%%%%%%%%%%%%%%%%%%%%%%%%%%%%%%%%%
%%%%%%%%%%%%%%%%%%%%%%%%%%%%%%%%%%%%%%%%%%%%%%%%%%%%%%%%%%%%%%%%%%%%%%%%%%%%%%%%%%%%%%%%%%%%%%%%%%%%%%%%%%%%%%%%%%%%%%%%%%%%%%%%%%%%%%%%%%%%%%%%%%%%%%%%%%%%%%%%
\section{Future Possibilities for the Pythagorean Formula}
\label{sec:future}

The Pythagorean Formula, in particular Pythag(1.83), has proven to be extremely robust. Many attempts have been made to improve upon it, with very little to show for a lot more work. For example, Luo-Miller \cite{LM} take into account park effects, and see essentially no improvement, while an effort to account for irrelevant runs in blowouts actually led to a worse predictor! Similar adjustments based on pitcher quality and others also do not lead to improvements.

%In our pursuit of a better method to predict wins from runs, we may need a completely different model. Well, how about a linear model? In \S\ref{sec:linear}, we show how a linear model to predict wins is actually the first-order multivariate Taylor expansion of the Pythagorean Formula! It is then unsurprising that it does not outperform Pythag(1.83), but having an easy to compute, linear estimate certainly has its benefits.

The shape parameter, $\gamma$, is equivalent to the exponent of the Pythagorean Formula in Miller's work. We have shown that having two different shape parameters, for RS and RA, can be useful for modelling. Future work should explore and see if a Pythagorean Formula with flexible exponents for RS and RA could outperform James' original formulation.

Another possibility for further research is to investigate further for which teams and seasons the method of moments outperforms Pythag(1.83). We expect an improvement when analysing teams that score and allow runs with very differently shaped distributions (such as the 2022 Washington Nationals, see Figure \ref{figure:WSNmoments}).

One can also explore if incorporating the third moment leads to any improvements. Just adding one more equation with the third moment could lead to issues, as we would now have three equations but only two unknowns. A possible resolution would be to let $\beta$ be a free parameter.

Finally, we discuss the potential of the Method of Moments to be applied to valuing players. In \S\ref{sec:applications}, we assess the value of a player by the runs they add or prevent to a team. However, it is plausible to suggest some players could significantly affect not just the mean runs of a team, but also the variance. For example, a carefree slugger and hardened walker might add a similar number of runs over the course of a season, but certainly one adds more variance than the other. By the Method of Moments, we can now account for that when analysing how many extra wins such a player might give a team.

\appendix

%%%%%%%%%%%%%%%%%%%%%%%%%%%%%%%%%%%%%%%%%%%%%%%%%%%%%%%%%%%%%%%%%%%%%%%%%%%%%%%%%%%%%%%%%%%%%%%%%%%%%%%%%%%%%%%%%%%%%%%%%%%%%%%%%%%%%%%%%%%%%%%
%%%%%%%%%%%%%%%%%%%%%%%%%%%%%%%%%%%%%%%%%%%%%%%%%%%%%%%%%%%%%%%%%%%%%%%%%%%%%%%%%%%%%%%%%%%%%%%%%%%%%%%%%%%%%%%%%%%%%%%%%%%%%%%%%%%%%%%%%%%%%%%
%%%%%%%%%%%%%%%%%%%%%%%%%%%%%%%%%%%%%%%%%%%%%%%%%%%%%%%%%%%%%%%%%%%%%%%%%%%%%%%%%%%%%%%%%%%%%%%%%%%%%%%%%%%%%%%%%%%%%%%%%%%%%%%%%%%%%%%%%%%%%%%
\section{Moments of The Weibull Distribution}
\label{sec:weibull}

The Weibull distribution is a continuous, three parameter distribution, with probability density function
\begin{equation}
    f(x;\alpha,\beta,\gamma) \ = \ \frac{\gamma}{\alpha}((x-\beta)/\alpha)^{\gamma -1}e^{-((x-\beta)/\alpha)^\gamma}
\end{equation}
for $x \geq \beta$. It is a very flexible distribution (see for example \cite{MABF} and the references therein); we saw this in Figure \ref{fig:pythagweibullgamma1to4}, where it can model many different one bump distributions by appropriately choosing values of the parameters.

One reason for its popularity is that straightforward integration suffices to obtain closed form expressions for its moments in terms of its parameters and the Gamma function $\Gamma(s)$; for the convenience of the reader we repeat the definition from \eqref{eq:gammadefn}: For $s\in\C$ with the real part of $s$ greater than $0$,
\begin{equation}
\Gamma(s)\ :=\ \int_0^\infty e^{-u} u^{s-1} {\rm d} u \ = \ \int_0^\infty e^{-u} u^s \frac{{\rm d} u}{u}.
\end{equation}

Denote the $k^{\rm th}$ moment of the Weibull distribution about $\beta$ by $m_k$ \footnote{For $X$ a random variable and $\beta\in \R$, the $k^{\rm th}$ moment around $\beta$ is $\mathbb{E}[(X-\beta)^k]$; thus if $X$ has density $p$ then $m_k = \int_{-\infty}^\infty (x-\beta)^k p(x) {\rm d}x$.}; we can easily find the mean and the variance of our distribution from $m_1$ and $m_2$, and thus do the more general calculation below and then specialize. The mean is just $$\E[X] \ = \ \E[X - \beta + \beta] \ = \ \E[X - \beta] + \E[\beta] \ = \ m_1 + \beta,$$ while the variance is \begin{eqnarray*} {\rm Var}(X) & \ = \ & \E[X^2] - \E[X]^2 \\ & = & \E[(X - \beta + \beta)^2] - \E[X - \beta + \beta]^2 \\ &=& \E[(X-\beta)^2 + 2 \beta (X - \beta) + \beta^2] - \left(\E[X - \beta] + \E[\beta]\right)^2 \\ &=& \left(\E[(X-\beta)^2] + 2 \beta \E[X - \beta] + \E[\beta^2]\right) - \left(m_1 + \beta\right)^2 \nonumber\\ &=& \left(m_2 + 2 \beta m_1 + \beta^2\right) - \left(m_1^2 + 2 \beta m_1 + \beta^2\right) \ = \ m_2 - m_1^2. \end{eqnarray*}

We have \begin{eqnarray*} m_k & \ = \ & \int_\beta^\infty (x-\beta)^k \cdot
\frac{\gamma}{\alpha} \left(\frac{x-\beta}{\alpha}\right)^{\gamma-1}
e^{-((x-\beta)/\alpha)^\gamma}{\rm d} x\nonumber\\ & = & \int_\beta^\infty  \alpha^k
\left(\frac{x-\beta}{\alpha}\right)^k \cdot \frac{\gamma}{\alpha}
\left(\frac{x-\beta}{\alpha}\right)^{\gamma-1}e^{-((x-\beta)/\alpha)^\gamma}{\rm d} x. \end{eqnarray*} Substituting $u =
\left(\frac{x-\beta}{\alpha}\right)^\gamma$, ${\rm d} u = \frac{\gamma}{\alpha}
\left(\frac{x-\beta}{\alpha}\right)^{\gamma-1}{\rm d} x$, we obtain \begin{eqnarray}
\label{eq:genweibmeans}
m_k & \ = \ & \int_0^\infty \alpha^k u^{k\gamma^{-1}} \cdot
e^{-u} {\rm d} u \nonumber\\ & = & \alpha^k \int_0^\infty e^{-u}
u^{1+k\gamma^{-1}} \frac{{\rm d} u}{u} \nonumber\\ & = & \alpha^k \ \Gamma(1+k\gamma^{-1}) \end{eqnarray}
by the definition of the Gamma function.

Denoting the mean by $\mu_{\alpha,\beta,\gamma}$ and the variance by $\sigma^2_{\alpha,\beta,\gamma}$, we find \begin{eqnarray*}
\mu_{\alpha,\beta,\gamma} &\ =\ & \alpha \Gamma\left(1+\gamma^{-1}\right) + \beta \nonumber\\ \sigma^2_{\alpha,\beta,\gamma}&\ =\ & \alpha^2 \Gamma\left(1+2\gamma^{-1}\right)  - \alpha^2 \Gamma\left(1+\gamma^{-1}\right)^2.
\end{eqnarray*}

%%%%%%%%%%%%%%%%%%%%%%%%
%%%%%%%%%%%%%%%%%%%%%%%%%%%%%%%%
%%%%%%%%%%%%%%%%%
\section{Deriving the Pythagorean Formula}\label{sec:millerRecap}

For completeness, we reproduce the argument of how James' Pythagorean prediction is a consequence of the assumptions that runs scored and allowed are independently drawn from Weibull distributions with the same parameter; see \cite{CGLMP,Mi}.

Let $X$ and $Y$ be independent random variables with Weibull distributions $(\alpha_\rs,\beta,\gamma)$ and $(\alpha_\ra,\beta,\gamma)$
respectively, where $X$ is the number of runs scored and $Y$ the
number of runs allowed per game. We wish to choose our parameters such that the means of our Weibull match the observed average runs scored and allowed, which we denote by RS and RA respectively.

We use \eqref{eq:genweibmeans}, and find \begin{eqnarray} \label{eq:pythagweibmeans} \alpha_\rs & \ = \ &
\frac{\rs-\beta}{\Gamma(1+\gamma^{-1})}, \ \ \ \ \ \ \ \alpha_\ra \ = \ \frac{\ra-\beta}{\Gamma(1+\gamma^{-1})}.\end{eqnarray}

The winning percentage is thus reduced to determining the probability that $X$ exceeds $Y$: \begin{eqnarray} & & \mbox{Prob}(X > Y) \ = \ \int_{x=\beta}^\infty \int_{y=\beta}^x f(x;\alpha_\rs,\beta,\gamma) f(y;\alpha_\ra,\beta,\gamma) {\rm d} y\; {\rm d} x \nonumber\\
& & = \
\int_{x=0}^\infty\frac{\gamma}{\alpha_\rs} \left(\frac{x}{\alpha_{RS}}\right)^{\gamma-1} e^{-(x/\alpha_\rs)^\gamma} \left[
\int_{y=0}^{x} \frac{\gamma}{\alpha_\ra}\left(\frac{y}{\alpha_{\ra}}\right)^{\gamma-1} e^{-
(y/\alpha_\ra)^\gamma} {\rm d} y \right] {\rm d} x  \nonumber\\ & & = \ \int_{x=0}^\infty\frac{\gamma}{\alpha_\rs}
\left(\frac{x}{\alpha_{RS}}\right)^{\gamma-1} e^{-(x/\alpha_\rs)^\gamma} \left[1
- e^{-(x/\alpha_{\ra})^\gamma}\right] {\rm d} x  \nonumber\\ & & = \ 1 -
\int_{x=0}^\infty\frac{\gamma}{\alpha_\rs}
\left(\frac{x}{\alpha_{RS}}\right)^{\gamma-1} e^{-(x/\alpha)^\gamma} {\rm d} x,\end{eqnarray}
letting \begin{equation} \frac1{\alpha^\gamma} \ = \ \frac1{\alpha_\rs^\gamma} +
\frac{1}{\alpha_\ra^\gamma} \ = \ \frac{\alpha_\rs^\gamma +
\alpha_\ra^\gamma}{\alpha_\rs^\gamma \alpha_\ra^\gamma}.\end{equation} Note we have reduced the problem to integrating a new Weibull with scale parameter\footnote{We often see similar expressions of how items combine; for example, in physics such combinations arise in center of mass calculations, or in adding resistors in parallel.} $\alpha$. Continuing, we have
\begin{eqnarray}\label{eq:derivweibpythag1} \mbox{Prob}(X
> Y) & \ = \ & 1 - \frac{\alpha^\gamma}{\alpha_\rs^\gamma} \int_{0}^\infty
\frac{\gamma}{\alpha}
\left(\frac{x}{\alpha}\right)^{\gamma-1} e^{(x/\alpha)^\gamma} {\rm d} x \nonumber\\
& = & 1 - \frac{\alpha^\gamma}{\alpha_\rs^\gamma} \nonumber\\ & = & 1 -
\frac1{\alpha_\rs^\gamma} \frac{\alpha_\rs^\gamma \alpha_\ra^\gamma}{\alpha_\rs^\gamma +
\alpha_\ra^\gamma} \nonumber\\ & = & \frac{\alpha_\rs^\gamma}{\alpha_\rs^\gamma +
\alpha_\ra^\gamma}.\end{eqnarray} Substituting in the relations for $\alpha_\rs$ and
$\alpha_\ra$ from (\ref{eq:pythagweibmeans}) gives \begin{eqnarray} \mbox{Prob}(X
> Y) & \ = \ &  \frac{(\rs-\beta)^\gamma}{(\rs-\beta)^\gamma + (\ra-\beta)^\gamma}, \end{eqnarray} returning the Pythagorean formula when $\beta = 0$ (which makes sense theoretically, as this is the minimum number of runs a team can score; as remarked above we often take $\beta=-1/2$ for binning purposes).

%%%%%%%%%%%%%%%%%%%%%%%%%%%%%%%%%%%%%%%%%%%%%%%%%%%%%%%%%%%%%%%%%%%%%%%%%%%%%%%%%%%%%%%%%%%%%%%%%%%%%%%%%%%%%%%%%%%%%%%%%%%%%%%%%%%%%%%%%%%%%%%
%%%%%%%%%%%%%%%%%%%%%%%%%%%%%%%%%%%%%%%%%%%%%%%%%%%%%%%%%%%%%%%%%%%%%%%%%%%%%%%%%%%%%%%%%%%%%%%%%%%%%%%%%%%%%%%%%%%%%%%%%%%%%%%%%%%%%%%%%%%%%%%
%%%%%%%%%%%%%%%%%%%%%%%%%%%%%%%%%%%%%%%%%%%%%%%%%%%%%%%%%%%%%%%%%%%%%%%%%%%%%%%%%%%%%%%%%%%%%%%%%%%%%%%%%%%%%%%%%%%%%%%%%%%%%%%%%%%%%%%%%%%%%%%
\section{Linearizing Pythagoras}\label{sec:linear}

The Pythagorean Won-Lost formula, see \eqref{eq:pythagWL}, was initially suggested by Bill James \cite{Ja} in the early 1980s. James originally proposed $\gamma$ to be 2 due to its ease of use, leading to the ``Pythagorean'' name. As remarked in the introduction, decades later in 2007 Miller \cite{Mi} offered the first statistical verification of the formula. By presuming that runs scored and runs allowed can be expressed as statistically independent Weibull distributions, he found that the probability of runs scored exceeding runs allowed yields Bill James' formula. Additionally, he found $\gamma$ to be approximately 1.82 by fitting the Weibull distributions to observed run production. Five years later, Dayaratna and Miller \cite{DM} derived a linear predictor for MLB teams' winning percentage by taking a first order approximation of Bill James' formula. They found the first order, multivariate Taylor series expansion of James' formula: \begin{equation} \frac{\#{\rm Wins}}{\#{\rm Games}} \ \approx \  .500 + \frac{\gamma}{4 \cdot {\rm R}_{{\rm ave}}} ({\rm RS}-{\rm RA}),\end{equation} where ${\rm R}_{{\rm ave}}$ is equal to the league-wide average runs scored over the course of a particular season. In doing so, they provided a justification for the simple linear predictor put forth by Jones and Tappin \cite{JT}, where the winning percentage is $.500 + \beta (\rs - \ra)$, and suggested that $\beta$ should be approximately $\gamma / (4 \cdot {\rm R}_{{\rm ave}})$, which is born out from seasonal data.

As this formula is easy to use and allows a quick estimate of the worth of increased run production or run prevention, we summarize its derivation. Our starting point is the second order Taylor series expansion of a function $f(x,y)$ about the point $(a,b)$:
\begin{eqnarray}
f(x,y) \ = \ f(a,b)&+&\frac{\partial f}{\partial x}\Big|_{(a,b)}(x-a) + \frac{\partial f}{\partial y}\Big|_{(a,b)}(y-b) + \frac{1}{2}\frac{\partial^2 f}{\partial x^2}\Big|_{(a,b)}(x-a)^2 \nonumber\\ & + & \frac{\partial^2 f}{\partial x\partial y}\Big|_{(a,b)}(x-a)(y-b) + \frac{1}{2}\frac{\partial^2 f}{\partial y^2}\Big|_{(a,b)}(y-b)^2 \nonumber \\ &+& \text{higher order terms.} \nonumber
\end{eqnarray}

\noindent Here, the higher order terms involve products of $(x-a)$ and $(y-b)$ to the third and higher powers, and thus are much smaller than the other terms when $x$ is close to $a$ and $y$ is close to $b$. A common technique in calculus is to replace a complicated function with a linear approximation, namely the tangent line in one dimension or the tangent plane in two; for us this means keeping just the constant and linear terms:
\begin{equation*}
f(x,y) \ \approx \ f(a,b) + \frac{\partial f}{\partial x}\Big|_{(a,b)}(x-a) + \frac{\partial f}{\partial y}\Big|_{(a,b)}(y-b).
\end{equation*}

\noindent Letting ${\rm R}_{{\rm ave}}$ denote the average number of runs scored in the league, we apply the above to James' Pythagorean estimate
\begin{equation*}
f(x,y) \ = \ \frac{x^\gamma} {x^\gamma + y^\gamma}
\end{equation*} and expand about the point $(a,b) = ({\rm R}_{{\rm ave}}, {\rm R}_{{\rm ave}})$. Taking $x={\rm RS}$ and $y={\rm RA}$ yields
\begin{eqnarray}
&f({\rm R}_{{\rm ave}},{\rm R}_{{\rm ave}}) \ = \ .500 \nonumber\\ &\frac{\partial f}{\partial x} \ = \ \frac{\gamma x^{\gamma-1} y^\gamma}{(x^\gamma+y^\gamma)^2} \ \Rightarrow \ \frac{\partial f}{\partial x}\Big|_{({\rm R}_{{\rm ave}},{\rm R}_{{\rm ave}})} \ = \ \frac{\gamma}{4\cdot{{\rm R}_{{\rm ave}}}} \nonumber\\ &\frac{\partial f}{\partial y} \ = \ -\frac{\gamma x^\gamma y^{\gamma-1}}{(x^\gamma+y^\gamma)^2} \ \Rightarrow \ \frac{\partial f}{\partial y}\Big|_{({\rm R}_{{\rm ave}},{\rm R}_{{\rm ave}})} \ = \ -\frac{\gamma}{4\cdot{{\rm R}_{{\rm ave}}}}. \nonumber
\end{eqnarray}

\noindent Noting that the predicted winning percentage is $f({\rm RS},{\rm RA})$, we see that the first order, multivariate Taylor series expansion about $({\rm RS},{\rm RA})$ implies
\begin{eqnarray*}
{\rm Winning\ Percentage} & \ \approx \ & .500 + \frac{\gamma}{4\cdot{{\rm R}_{{\rm ave}}}}({\rm RS}-{\rm R}_{{\rm ave}}) - \frac{\gamma}{4\cdot{{\rm R}_{{\rm ave}}}}({\rm RA}-{\rm R}_{{\rm ave}}) \nonumber\\ &= & .500 + \frac{\gamma}{4\cdot{{\rm R}_{{\rm ave}}}}({\rm RS}-{\rm RA}).
\end{eqnarray*}

The slope coefficient $\gamma / (4 \cdot {\rm R}_{{\rm ave}})$ can be easily computed using standard linear regression techniques. Thus, the value of $\gamma$ can be directly estimated by multiplying the slope coefficient by $4\cdot{{\rm R}_{{\rm ave}}}$. Note of course that this analysis crucially depends on the shape of James' Pythagorean predictor; a different function would have different partial derivatives, leading to another estimator. For other candidates, see the work of Hammond, Johnson and Miller \cite{HJM}.

With Bill James's original formula, the use of squared powers renders expected won-lost percentages easy to compute on a calculator.  Although improvements to statistical computing in the decades since have certainly made dealing with powers of $\gamma$ (including estimates around 1.8 as estimated in Miller \cite{Mi}) even easier, it is nevertheless useful to have good approximations that are quick and easy to use and give a ``ballpark'' sense of what is going on.  The linear approximation presented above does precisely this by offering a much simpler method suitable for a non-technical audience, and can be easily implemented in commonly used programs such as Microsoft Excel or Google Sheets.

%%%%%% COMPARE METHODS HERE %%%%%%
\begin{table}[h]
\begin{center}
\resizebox{\textwidth}{!}{%
\begin{tabular}{lrrrrrrrrrrrr}
\hline
 {\rm Method} & \ &   {\rm '22 Avg} & \ & {\rm '12 Avg} & \ &   {\rm '22 Standard Deviation} & \ & {\rm '12 Standard Deviation} & \ &  {\rm '22 Median} & \ &  {\rm '12 Median} \\ \hline
 {\rm Moments} & \ &   2.84 & \ & 3.10 & \ &    2.02 & \ & 2.28 & \ &  2.23 & \ & 2.56 \\
 {\rm Pythag(1.83)} & \ &   2.63 & \ & 2.94 & \ &   1.99 & \ & 2.43 & \ &   2.18 & \ & 2.19  \\
 {\rm Least Squares with $\gamma_\rs = \gamma_\ra$} & \ &   2.47 & \ & 3.66 & \ &   2.30 & \ & 2.79 & \ & 1.63 & \ & 2.80  \\
 % {\rm Moments with $\gamma_\rs = \gamma_\ra$} & \ &  6.60  & \ & 8.39 & \ & 5.44  & \ & 6.48 & \ & 4.99 & \ & 6.87  \\
 {\rm Least Squares with $\gamma_\rs,\gamma_\ra$ free} & \ & 2.72   & \ & 3.42 & \ & 2.01   & \ & 2.32 & \ & 2.08 & \ & 3.36  \\
 {\rm Linear Predictor} & \ &   2.56 & \ & 2.76 & \ &   3.21 & \ & 3.76 & \ & 2.30 & \ & 1.70  \\
\hline
\end{tabular}%
}
\caption{\label{table:comparepredict} Comparing the mean, standard deviation, and median of the absolute difference between the observed wins and the predicted wins by different methods in the 2022 and 2012 MLB seasons.}
\end{center}
\end{table}
%%%%

Using data from the 2012 and 2022 MLB seasons, we estimate win percentage (WP) by the linear regression model: $\text{WP}=.500+\beta(\RS-\RA)$. We find the slope $\beta$, for 2012 and 2022 respectively, to be 0.11236 and 0.09835. From here, we can simply use Dayaratna and Miller's \cite{DM} first order approximation to estimate $\gamma$. We find $\gamma$ to be approximately 1.94 for the 2012 season and 1.69 for the 2022 season by multiplying each season's slope by $4\cdot\Rave$.

Assuming a significance level of $\alpha=.05$, the slope $\beta$ for both the 2012 and 2022 seasons are highly significant with $p<0.01$. Even after adjusting for multiple comparisons via Bonferroni corrections to a threshold of $\alpha=.025$, both slopes are still significant. Additionally, the high R-squared statistics indicate the model's strong fit to the data given; in 2012, 89.81\% and in 2022, 94.82\% of the variability in winning percentage can be explained by the average run differential per game ($\RS-\RA$). Finding a 95\% confidence interval for the slope $\beta$ is useful because we can then calculate an interval for $\gamma$ by multiplying by $4\cdot\Rave$. Using a t-distribution with 28 degrees of freedom, we get a confidence interval of (0.08944, 0.10725) for the 2012 season's slope and (0.09770, 0.12701) for the 2022 season's slope. Translating these intervals into $\gamma$ yields confidence intervals of approximately (1.69, 2.20) for the 2012 season's $\gamma$ and (1.53, 1.84) for the 2022 season's $\gamma$.

Using our linear estimate of $\gamma$ for each respective season as the exponent in the Pythagorean Won-Lost formula: $\rs^\gamma / (\rs^\gamma + \ra^\gamma)$, we can compare our linear predictor's accuracy to other methods. Table \ref{table:comparepredict} above shows the mean, standard deviation, and median of the absolute difference between the observed wins and predicted wins. The linear model's low mean and median absolute difference are on par, if not marginally better than other methods. However, it's important to note the higher variability in the linear model's predictions, as shown by its notably larger standard deviation of absolute differences. As a result, the linear approximation has less predictive power than other methods, but remains a reasonable option for the average fan due to its ease of use and understanding.

%%%%%%%%%%%%%%%%%%%%%%%%%%%%%%%%%%%%%%%%%%%%%%%%%%%%%%%%%%%%%%%%%%%%%%%%%%%%%%%%%%%%%%%%%%%%%%%%%%%%%%%%%%%%%%%%%%%%%%%%%%%%%%%%%%%%%%%%%%%%%%%
%%%%%%%%%%%%%%%%%%%%%%%%%%%%%%%%%%%%%%%%%%%%%%%%%%%%%%%%%%%%%%%%%%%%%%%%%%%%%%%%%%%%%%%%%%%%%%%%%%%%%%%%%%%%%%%%%%%%%%%%%%%%%%%%%%%%%%%%%%%%%%%
%%%%%%%%%%%%%%%%%%%%%%%%%%%%%%%%%%%%%%%%%%%%%%%%%%%%%%%%%%%%%%%%%%%%%%%%%%%%%%%%%%%%%%%%%%%%%%%%%%%%%%%%%%%%%%%%%%%%%%%%%%%%%%%%%%%%%%%%%%%%%%%

\ \\

\end{document}